\documentstyle[amscd,amssymb,verbatim,12pt]{amsart}
\theoremstyle{plain}

\theoremstyle{definition}

\numberwithin{equation}{section}

\def\dspace{\baselineskip=0.3 in}
\begin{document}
\dspace
\title[Quantum Driven Bounce.................]{Quantum Driven Bounce Of The Future Universe  }

\author[S.K.Srivastava]%
        {    }

\maketitle

\centerline{\bf S.K.Srivastava }

\centerline{ Department of Mathematics, North Eastern Hill University,}

\centerline{ NEHU Campus,Shillong - 793022 ( INDIA ) }

\centerline{e-mail:srivastava@@nehu.ac.in ; sushil@@iucaa.ernet.in }

\vspace{2cm}

\centerline{\bf Abstract}

 It is demonstrated that due to back-reaction of quantum effects, expansion of
 the universe stops at its maximum and takes a turn around. Later on, it
 contracts to a very small size in finite future time. This phenomenon is
 followed by a ``bounce'' with re-birth of an exponentially expanding
 non-singular universe.

PACS no. 98.80.Cq

\newpage

Astrophysical data \cite{ar, ag} show cosmic acceleration in the
current universe, which is fuelled by dark energy (DE) violating strong or
weak energy condition (SEC or WEC). If DE mimics quintessence, the equation of
state parameter (EOSP) ${\rm w} > -1$ and SEC is violated . In the case of
phantom (super-quintessence) DE, ${\rm w} < -1$ and WEC is violated.  It is
found that when w$ < - 1$, DE leads the cosmological model having ``big- rip singularity'' and ending
in  cosmic doomsday with infinite energy density \cite{ rrc}. A comprehensive
review on DE and accelerating universe is available in \cite{jc}. In contrast to quintessence model, phantom model
shows growing curvature with time and in models with ``big-rip singularity'' at finite time $t_s$ with $t$ being
the cosmic time, curvature
invariants (containing high powers of curvature) grow very strong, when $|t_s
- t| $ is sufficiently small. Quantum gravity suggests that quantum corrections
to matter fields like scalars, spinors and vectors, in curved space-time, depend upon curvature. It
implies dominance of quantum gavity near $t= t_s$. 
 It is analogous to the early universe model, where
curvature invariants are very strong and cause effective role of quantum corrections. This fact
was pointed out in  \cite{ee, sn}, where it is demonstrated that singularity
may be avoided or made milder using quantum corrections near
{\em future singularity time} $t_s$.

Like \cite{sn}, here also, scale factor of the universe is assumed in the
beginning. Further, it is probed what type of conformal scalar (quintessence
or phantom) support this scale factor (the assumed one) for being the solution
of Friedmann equations. Like \cite{ee,sn}, here also, quantum gravity effects
are investigated when $t$ approaches $t_s$. In \cite{sn}, a possible escape
from cosmic doomsday at $t=t_s$ is discussed using quantum corrections. Here,
the approach is more elegant and a cosmic scenario is obtained, where
expansion stops at $t=t_m$ very near to  $t_s$. As a result, it takes a turn
around and contracts for a small period $(t_s - t_m)$. Subsequently, at
$t=t_s$, cosmic bounce takes place and universe expands exponentially for
$t>t_s$. This scenario emerges due to quantum effects during the period $(t_s
- t_m)$. Thus, here, singularity is not made milder , but it is completely avoided in
contrast to results in \cite{ee, sn}. In the present paper, this scenario is
obtained for (i) general relativity (GR) model, (ii)
Randall- Sundrum II (RSII) model and (iii) Gauuss-Bonnet(GB) model.

 It is
found that the scale factor reaches its maximum in finite time $t_m$, when
scalar is phantom (super-quintessence) in GR and RSII models. But, in
GB-model, similar situation is obtained for both quintessence conformal scalar
and conformal phantom subject to different conditions. In \cite{msak} also,
conformal phantom scalar, in FRW universe, had been considered in a different
context. Here, an escape from `future singularity' in phantom model is
explored using quantum gravity effects of conformal phantom scalars, whereas, in \cite{msak},
non-integrability of hamiltonian system of these fields is discussed using Ziglin theorems on
integrability. Natural units $({\hbar} = c = 1)$ are used with $\hbar$ and $c$
having their usual meaning. GeV is the fundamental unit and cosmic time $t$ is
measured in GeV$^{-1}$.

\bigskip
\noindent \underline{\bf General Relativity model of the future universe}

\noindent {\bf (a) Classical approach}

\smallskip
According to experimental probes \cite{adm}, our universe is spatially flat with growing scale factor $a(t)$, given by the distance function

$$ dS^2 = dt^2 - a^2(t)[ dx^2 + dy^2 + dz^2 ]  \eqno(1)$$
such that ${\ddot a} = \frac{d^2a}{dt^2} > 0.$ 

DE density $\rho$ and pressure $p$ are given by Friedmann equations

$$ H^2 = \Big(\frac{\dot a}{a}\Big)^2 = \frac{8 \pi G}{3} \rho  \eqno(2a)$$
and
$$\frac{\ddot a}{a} = - \frac{4 \pi G}{3}( \rho + 3 p ),
 \eqno(2b)$$
where $G = M_P^{-2}$ ($M_P$ is the Planck's mass).

Here, source of the dark energy is  conformal scalar  $\phi(x)$ given by
the action 

$$ S_{\phi} = \int {d^4x} \sqrt{-g} \frac{1}{2}\Big[\omega^2
g^{\mu\nu}{\partial}_{\mu}\phi {\partial}_{\nu}\phi - \frac{1}{6} R
\phi^2 \Big], \eqno(3a)$$
where $\omega^2 = \pm 1$ for quintessence and phantom (super-quintessence) scalars respectively.

This action yields the Klein-Gordon equation
$$ \omega^2{\ddot \phi} + 3\omega^2  H {\dot \phi} + ({\dot H} + 2 H^2) \phi =
0, \eqno(3b)$$ 
where $R = 6 ({\dot H} + 2 H^2)$ and $\phi = \phi(t)$ due to homogeneity of
the space-time (1).

The energy density $\rho$ and pressure $p$ are obtined from the action (3a) as

$$ \rho = \frac{1}{2}\omega^2 {\dot \phi}^2 + H \phi {\dot \phi} +
\frac{1}{2} H^2 \phi^2 \eqno(3c)$$
and
$$ p = \frac{1}{2}\omega^2 {\dot \phi}^2 - \frac{1}{3} (\phi {\ddot \phi} +
{\dot \phi}^2 + 3 H \phi {\dot \phi}) - \frac{1}{6} (2 {\dot H} + 3 H^2)
\phi^2. \eqno(3d)$$

The conservation equation is 
$$ {\dot \rho} + 3 H (\rho + p) = 0 . \eqno(3e)$$

It is noted that the Klein-Gordon equation (3b) can be derived connecting eqs.(3c), (3d) and
(3e). So, eqs.(3b) and (3e) are not independent implying that it is enough to
use either of these two equations. Now, we are left with three
equations (2a), (2b) and (3b) as well as (3c) and (3d) for $\rho$ and $p$
respectively. Connecting (2a) and (3c), it is obtained that

$$ {\dot \phi} = H [- \omega^2 \phi  \pm \sqrt{\frac{3 \omega^2}{4 \pi G} +
  \phi^2 (1 - \omega^2)}].
\eqno(4a)$$

This equation shows that when $\omega^2 = -1, \phi^2 \ge {3}/{8 \pi G}$.
 Using eqs.(3b), (3c), (3d) and (4a) in
eq.(2b), it is obtained that

$$ \aleph \phi^4 + 2 \vartheta \phi^2 + \varkappa = 0 \eqno(4b)$$
with
$$\aleph = [(6 \omega^2 - 5) H^2 + (\omega^2 - 1){\dot H}]^2 - 4(1 -
\omega^2)(\frac{3}{2}\omega^2 - 1) H^2, \eqno(4c)$$

$$ 2 \vartheta = \frac{3}{2\pi G}[{\dot H} + (3 - \omega^2) H^2](6 \omega^2 - 5)
H^2 + (\omega^2 - 1){\dot H}] - \frac{3}{\pi G}\omega^2(\frac{3}{2}\omega^2 -
1) H^2  \eqno(4d)$$
and
$$\varkappa = \Big[\frac{3}{4 \pi G} \Big({\dot H} + (3 - \omega^2) H^2 \Big)
\Big]^2 . \eqno(4e)$$
Eq.(4b) is obtained writing ${\ddot a}/a = {\dot H} + H^2 $ as well as using
${\dot \phi}, {\ddot \phi}$ in terms of $\phi, H$ and ${\dot H}$. This
equation yields
$$ \phi^2 = \frac{ - \vartheta  \pm \sqrt{\vartheta^2 - \aleph \varkappa}}{\aleph}. \eqno(4f)$$

Thus it is obtained that an $a(t)$ will satisfy eqs.(2a) and (2b), if
      conformal scalar $\phi$ obeys eqs.(4a) and (4b)\cite{vf}. Now, we can
      use conditions (4a) and (4b) to examine an $a(t)$ for being solutions of
      Friedmann equations (2a) and (2b) with conformal scalar $\phi$ as a
      source of DE.

For the time period $t_0 \le t \le t_m,$ the form of $a(t)$ is assumed to be
$$ a(t) = A + B (M_P t )^r + |M_P(t_s -t)|^{-q}, \eqno(5a)$$
where $r > 1$ to get ${\ddot a(t)} > 0$, $q > 0, A$ and $B$ are arbitrary
constants. The classical approach shows that the scale factor (5a) has big-rip
singularity at $t=t_s$ \cite{mpd}. In the following section, it is shown that an escape
from this catastrophic situation is possible using quantum gravity effects. 

Now, we explore whether $\phi$ mimics quintessence or phantom, when $a(t)$
(given by eq.(5a)) satisfies eqs.(2a) and (2b) as well as acquires its maximum
at time $t = t_m$.  It turns out to explore $\phi$, when $a(t)$
satisfies eqs.(4a) and (4b) for $t_0 \le t \le t_m$ with $a(t_m)$ being the maximum.  With $a_0 = a(t_0) (t_0$ being the present time) $A$ is obtained as

$$ A = a_0 - B(M_P t_0 )^r -  \Big|M_P (t_s -t_0) \Big|^{-q} . \eqno(5b)$$

$a(t)$, given by eq.(5a), expands to its maximum by the time $t = t_m,$ if
$$ {\dot a(t_m)} = r B M_P^r t_m^{(r-1)} + q M_P^{-q} |t_s - t_m|^{(-q -1)} =
0$$
implying
$$ B = - \frac{q M_P^{(-q-r)}|t_s - t_m|^{(-q -1)}}{r t_m^{(r-1)}}.
\eqno(6)$$

Thus ,$a(t)$ from eq.(5a), looks like
$$a(t) = a_0 - \frac{q}{r} \frac{M_P^{-q}}{t_m^{(r-1)}}|t_s - t_m|^{(-q-1)}
(t^r - t^r_0) + M_P^{-q}|t_s - t|^{-q} -  M_P^{-q}|t_s - t_0|^{-q} \eqno(7)$$
for $t_0 \le t \le t_m$.

Moreover,
$${\dot a(t)} = q M_P^{-q} |t_s - t_m|^{(-q -1)} \Big[ - \Big(t/t_m
\Big)^{(r-1)} + \Big(|t_s - t|/|t_s - t_m| \Big)^{(-q - 1)} \Big] 
\eqno(8)$$
and
$${\ddot a(t)} = q M_P^{-q} |t_s - t_m|^{(-q -1)} \Big[ - \frac{(r-1)}{t}\Big(t/t_m
\Big)^{(r-1)}$$
$$ + \frac{(q+1)}{|t_s - t|} \Big(|t_s - t|/|t_s - t_m| \Big)^{(-q - 1)} \Big]
. \eqno(9)$$ 
These equations show ${\dot a(t)} >0$ and ${\ddot a(t)} > 0$ for  $t_0 < t <
t_m$ and $q > 0, r > 1$ as 
$$ \frac{1}{|t_s - t|} > \frac{1}{t} . \eqno(10)$$
Eqs.(8) and (9) yield that, at $t = t_m, H = 0$ and ${\dot H} \ne 0.$

Now, we have two cases (i) $\omega^2 = 1$ and (ii) $\omega^2 = -1$. Eq.(4f)
yields $\phi^2$ being {\em indeterminate} for $\omega^2 = 1$, when $H = 0$ and
${\dot H} \ne 0$ at $t = t_m$ . But, for $\omega^2 = -1$ 
$$ \phi^2(t_m) = \frac{3}{8 \pi G} \eqno(11)$$ 
for $H(t_m) = 0$ and ${\dot H(t_m)} \ne 0.$

It means that the scale factor $a(t)$, given by eqs.(5a), (5b) and (6) or
eq.(7), is not a solution of eqs.(2a) and (2b) for the time period 
$t_0 \le t \le t_m$, if $\phi$ is a quintessence scalar, but it is a solution
of eqs.(2a) and (2b) for the time period 
$t_0 \le t \le t_m$ for $\phi$ being a phantom scalar characterized by
$\omega^2 = -1$.

This analysis suggests that if universe is dominated by quintessence dark
energy, expansion will not stop at its maximum. So, in this model (GR model),
only the case $\omega^2 = -1$ is investigated.

\newpage

\noindent \underline{\bf Quantum effects and bounce near $t_s$}

\smallskip

In the scenario discussed above, ${\dot a(t_m)} = 0$, which means that either
  $\rho$ in eq.(2a) vanishes at $t = t_m$ or there should be a term $\rho_{\rm corr}$ on the r.h.s. of the
  equation (2a), which remains ineffective during the time interval $t_0 < t <
  t_m$, but at $t = t_m$ it makes the effective energy density $\rho_{\rm eff}
  = \rho + \rho_{\rm corr} = 0.$

Using eqs.(4a), (5a,b) and (6) in eq.(3c), it is found that $\rho$ increases
with time during the interval $t_0 \le t \le t_m$. So, only second possibility
may be valid as $\rho(t_m) \ne 0.$ This argument modifies the Friedmann
equation (2a) as
$$ \Big(\frac{\dot a}{a} \Big)^2 = \frac{8 \pi G}{3}(\rho + \rho_{\rm corr})         \eqno(12a)$$
and eq.(2b) as
$$\frac{\ddot a}{a} = - \frac{4 \pi G}{3}( \rho + \rho_{\rm corr} + 3 p + 3 p_{\rm corr} ),
 \eqno(12b)$$
such that for $t < t_m, \rho_{\rm corr} \approx 0$ and at $t = t_m,  \rho = -
 \rho_{\rm corr}.$ In what follows, we explore $\rho_{\rm corr}$ satisfying
 these requirements. 

As
it is found in eq.(10) that $t_m$ also satisfies the inequality $|t_s - t_m| <
1$. So, when $t_s \gtrsim t \ge t_m$, curvature becomes very strong and quantum gravity
becomes effective \cite{ee,sn,ndb}. Conformal anamoly for scalars, due to
 quantum effects, 
yield correction to $\rho$ and $p$ as
 
$$\rho_A = \frac{N }{60 (4\pi)^2} H^4 - \frac{N }{180 (4\pi)^2} \{ - 6 H {\ddot H} - 18 H^2 {\dot H} + 3 {\dot H}^2 \}  \eqno(13)$$
and
$$p_A = -\frac{N }{360 (4\pi)^2} ( 6 H^4 + 8 H^2 {\dot H} ) - \frac{N }{180 (4\pi)^2} \{ 2 \frac{d{\ddot H}}{dt} + 12 H {\dot H} + 18 H^2 {\dot H} + 9 {\dot H}^2 \}, \eqno(14)$$
where $N$ is the number of scalars.

Now, connecting eqs.(5a), (5b), (13) and (14), for $t_0 < t < t_m,$ it is obtained
that
$$ \rho_A = - \frac{N q M_P^{-2q}}{60 (4\pi)^2} |t_s - t|^{(-2q -4)} \Big[(q
+1) $$
$$ - (r-1) \frac{|t_s - t_m|^{(-q - 1)}}{t} \Big(t/t_m \Big)^{(r-1)} |t_s
  -t|{(q +2)} \Big]^2 . \eqno(15)$$
This equation shows that for $t < t_m, \rho_A \approx 0.$ But at $t = t_m,$
$$ \rho_A (t_m) = - \frac{N q M_P^{-2q}}{60 (4\pi)^2} |t_s - t_m|^{(-2q -4)} \Big[(q
+1) $$
$$ - (r-1) \frac{|t_s - t_m|^{-q }}{t_m}(q +2) \Big]^2 . \eqno(16)$$
Thus, we find that $\rho_A$ satisfies the criteria required by $\rho_{\rm
  corr}$. So, it is set that $\rho_{\rm corr} = \rho_A$.

As $H(t_m) = 0$,  eq.(13) yield $\rho(t_m) +  \rho_A (t_m) = 0,$ so
$$ \rho(t_m) =  \frac{N q M_P^{-2q}}{60 (4\pi)^2} |t_s - t_m|^{(-2q -4)} \Big[(q
+1) $$
$$ - (r-1) \frac{|t_s - t_m|^{-q }}{t_m}(q +2) \Big]^2 . \eqno(17)$$

Thus it is obtained that DE density grows with time . Moreover, the correction
term which is negligible for $t < t_m$ becomes very strong at $t = t_m$. As
$ \rho_A < 0, \rho_{\rm eff}(t_m) = \rho(t_m) + \rho_A(t_m) = 0$ , but for $t <
t_m,$  $\rho_{\rm eff}(t_m) = \rho.$ So, it is quantum effect, which stops
expansion of the universe at $t = t_m$.

For the modified Friedmann equations (12a) and (12b), solution is taken as
$$ a(t) = exp [ B^{\prime} (M_P t)^{\alpha} + C^{\prime} M_P^{\beta} (t -
t_m)^{\beta} + M_P^{\gamma} |t_s - t|^{\gamma} ] , \eqno(18a)$$
when $t > t_m$. 
Moreover, $a(t)$ should be continuous at $t = t_m,$ which is given by the
condition $ \lim_{\epsilon \to 0} {\dot a(t_m + \epsilon)} =  \lim_{\epsilon
  \to 0} {\dot a(t_m - \epsilon)} .$ Using this condition for $a(t)$, given
by eq.(5a) (when $t < t_m$) and $a(t)$, given by eq.(18a) (when $t > t_m$), it
is obtained that
$$ \alpha B^{\prime} M_P^{\alpha} t_m^{(\alpha -1)} - \gamma  M_P^{\gamma} |t_s
- t_m|^{(\gamma - 1)} = 0, $$
 yielding
$$ B^{\prime} = \frac{\gamma  M_P^{\gamma} |t_s
- t_m|^{(\gamma - 1)}}{\alpha M_P^{\alpha} t_m^{(\alpha -1)}}.  \eqno(18b)$$
 
For $t > t_m,$ eq.(18a) yields
\begin{eqnarray*}
H &=&  \alpha B^{\prime} M_P^{\alpha}t^{(\alpha -1)} + \beta
  C^{\prime} M_P^{\beta}(t - t_m)^{(\beta - 1)} - \gamma M_P^{ \gamma}
  |t_s - t|^{(\gamma -1)}, \\ {\dot H} &=& \alpha (\alpha -1)B^{\prime}
  M_P^{\alpha}t^{(\alpha -2)} + \beta (\beta - 1)
  C^{\prime} M_P^{\beta}(t - t_m)^{(\beta - 2)} \\ &&+ \gamma (\gamma -1)M_P^{
  \gamma}|t_s - t|^{(\gamma -2)}, \\   {\ddot H} &=& \alpha (\alpha -1)(\alpha -2)B^{\prime}
  M_P^{\alpha}t^{(\alpha -3)} + \beta (\beta - 1)(\beta - 2)
  C^{\prime} M_P^{\beta}(t - t_m)^{(\beta - 3)} \\ &&- \gamma (\gamma -1)(\gamma -2)M_P^{
  \gamma}|t_s - t|^{(\gamma -3)} ,  \\ {\dddot H}&=& \alpha (\alpha -1)(\alpha -2)(\alpha -3)B^{\prime}
  M_P^{\alpha}t^{(\alpha -4)} + \beta (\beta - 1)(\beta - 2) \\ && \times(\beta - 3)
  C^{\prime} M_P^{\beta}(t - t_m)^{(\beta - 4)} + \gamma (\gamma -1)(\gamma -2)(\gamma -3)M_P^{
  \gamma}|t_s - t|^{(\gamma -4)}.
\end{eqnarray*}
$$ \eqno(19a,b,c,d)$$

If $a(t)$, given by eq.(18a), satisfies eqs.(12a) and (12b), $\rho$ and $p$
are obtained as
\begin{eqnarray*}
 \rho &=& \frac{3 H^2}{8 \pi G} - \rho_A \\ &=& \frac{3 H^2}{8 \pi G} -
 \frac{N}{180 (4 \pi)^2} [ 3 H^4 + 6 H {\ddot H} + 18 H^2 {\dot H} - 3 {\dot
 H}^2 ]\\ p &=& - \frac{1}{4 \pi G} ({\dot H} + \frac{3}{2} H^2) +
 \frac{N}{360 (4 \pi)^2} ( 6 H^4 + 8 H^2 {\dot H} ) \\&& + \frac{N}{180 (4
 \pi)^2} [ 2 {\dddot H} + 12 H {\dot H} + 18 H^2 {\dot H} + 9 {\dot H}^2 ],
 \end{eqnarray*}  
$$ \eqno(20a,b)$$
where $ H, {\dot H}, {\ddot H}$ and ${\dddot H}$ are given by eqs.(19a,b,c,d).

Eqs.(19 a,b,c,d) and (20a,b) show that $\rho$ and $p$ can be finite at $t =
t_s,$ if $\gamma \ge 4$. So, here onwards, $\gamma = 4$ is set in equations.
Moreover , without any harm to physics, $\alpha$ and $\beta$ are
also taken as $\alpha = \beta = 3$. 

Thus $a(t)$, given by eq.(18a),is obtained as
$$ a(t) = exp [ B^{\prime}(M_Pt)^3 + C^{\prime} M_P^3 (t - t_m)^3 + M_P^4 |t_s
- t|^4 ], \eqno(21a)$$
where $t > t_m$.  If $a(t)$, given by eq.(22a), acquires its minimum $a_s$ at $t = t_s, {\dot
  a(t_s)} = 0$ yielding
$$  C^{\prime} = - \frac{\alpha}{\beta} M_P^{(\alpha - \beta)}
\frac{t_s^2} {(t_s - t_m)^2} B^{\prime}.  \eqno(21b)$$
Moreover, eq.(18b) is re-written as
$$ B^{\prime} = \frac{4  M_P |t_s
- t_m|^2}{3 t_m^2}.  \eqno(21c)$$
 
Use of $\alpha = \beta = 3$ and $\gamma = 4$, in eq.(19a), yields $H < 0$
implying ${\dot a(t)} < o$ during the period $t_m < t < t_s$. It shows
contraction of the universe during this period. For $t > t_s,$ eq.(19a) yields
$$ H =  4 M_P^4|t_s - t_m|^3 \Big(\frac{t_s}{t_m } \Big)^2 \Big[ \Big(\frac{t}{t_m } \Big)^2 -
\Big(\frac{[t - t_m]}{[t_s - t_m]} \Big)^2 + \Big(\frac{t_s}{t_m } \Big)^2
\Big(\frac{|t_s - t|}{|t_s - t_m |} \Big)^3 \Big]> 0  \eqno(22)$$
showing expansion . At $t = t_s,$ eq.(21a) yields 
$$ a(t_s) = exp \Big[\frac{4}{3} M_P^4 \frac{t_s^2}{t_m}(t_s - t_m)^3 \Big]. \eqno(24)$$         

Thus, it is obtained that, during the period $t_m < t < t_s$, universe will
contract and will acquire its minimum at $t=t_s$. Later on, it will expand for
$t > t_s$ showing a bounce at  $t = t_s.$

 As $H(t_s) = {\dot a(t_s)}/a(t_s) = 0$, 
eqs.(13 ) and (19 a,b,c,d) yield 
$$ \rho(t_s)= \frac{N}{15 \pi^2} M_P^8 (t_s - t_m)^4 \Big(\frac{t_s}{t_m }\Big)^2
\eqno(24a)$$
and
$$ p(t_s) = \frac{N}{60 \pi^2} M_P^4 + \frac{2 M_P^6}{\pi} (t_s - t_m)^2
\Big(\frac{t_s}{t_m } \Big) + \frac{ N}{4800 \pi^4} M_P^8 (t_s - t_m)^4
\Big(\frac{t_s}{t_m } \Big)^2. \eqno(24b)$$

\bigskip

\noindent \underline{\bf  RS II brane-world model}

\noindent{\bf Classical approach}

\smallskip

It is mentioned above that curvature invariants in future universe, dominated
by phantom dark energy, get stronger with increasing time. So , relevance of high energy physics increases
\cite{mg}. With this view, above type of analysis is done in brane-world also.
Here, RSII model is considered due to its simplicity and geometric
appeal. Moreover, it is more relevant to cosmology compared to RSI
model. Modified version of eqs.(2a,b), for brane-gravity,  are written as
\cite{sn,rm}

$$ H^2 = \frac{8 \pi G}{3} \rho_{RS} \Big[ 1 + \frac{\rho_{RS}}{2 \lambda} \Big] \eqno(25a)$$
and

$$\frac{\ddot a}{a} = - \frac{4 \pi G}{3}\Big[ \rho_{RS} + 3 p_{RS}  +  \frac{1}{\lambda} \rho_{RS}(2 \rho_{RS} + 3 p_{RS} ) \Big],
 \eqno(25b)$$
where $\lambda$ is the brane-tension.

Eq.(25a) yields

$$\rho_{RS} = \lambda \Big[ - 1 \pm \sqrt{1 + \frac{3 M_P^2 H^2}{16 \pi
    \lambda}   } \Big] .\eqno(26)$$

Eqs.(3c) and (26) yield

$$ {\dot \phi} = \omega^2 \Big[ - H \phi  \pm \sqrt{(1 - \omega^2) H^2 \phi^2
      + 2 \omega^2\lambda \Big[ - 1 \pm \sqrt{1 + \frac{3
      M_P^2 H^2}{4 \pi \lambda}  } \Big]}\Big] , \eqno(27)$$

In this model also, matter field is given by the same scalar field $\phi,$ so
for $\rho_{RS}$ and $p_{RS}$ also eqs.(3c) and (3d) are used in eq.(25b). As a
result,  it is obtained that

$$ - \frac{1}{4 \pi G} ({\dot H} + 3 H^2) = \Big\{\Big(\frac{\omega^2}{2} -
\frac{1}{3} \Big) \Big[- 1 \pm \sqrt{1 + \frac{3 M_P^2 H^2}{16 \pi
    \lambda}   } \Big] - \frac{1}{3} \Big\} \times \Big[ - H \phi $$
$$ \pm \sqrt{(1 - \omega^2) H^2 \phi^2 + 2 \lambda \omega^2 \Big[- 1 \pm\sqrt{1 +
    \frac{3 M_P^2 H^2}{16 \pi \lambda} } \Big] } \Big]^2 $$ 
$$ - \omega^2
\Big[- H^2 \phi^2 \pm \sqrt{(1 - \omega^2) H^2 \phi^2 + 2 \lambda \omega^2
  \Big[- 1 \pm\sqrt{1 + \frac{3 M_P^2 H^2}{16 \pi \lambda} } \Big] } \Big]$$
$$ - \phi^2 \Big[ - \frac{\dot H}{3} + H^2 + \Big\{\frac{\omega^2}{3} ({\dot
  H} + H^2) - \Big(\frac{\dot H}{3} + \frac{H^2}{2} \Big) \Big\} \Big[[- 1
\pm\sqrt{1 + \frac{3 M_P^2 H^2}{16 \pi \lambda} } \Big] \Big] .  \eqno(28)$$

$a(t)$, given by eq.(5a), acquires its maximum at $t=t_m, $ so ${\dot a}(t_m)
= 0$, but ${\ddot a}(t_m) \ne 0.$ It yields ${\dot H}(t_m) \ne 0.$ In case,
$\omega^2 = 1,$ eq.(28) implies that either $\dot H$ vanishes or it is
complex. So, like GR-based model , $a(t)$ does not satisfy modified Friedmann
equations (25a,b) for RSII model too, when $\omega^2 = 1$ . But, in case of $\omega^2 = - 1,$ $a(t)$,
given by eq.(5a) satisfies eqs.(25a,b) and at $t=t_m, H(t_m) = 0$ but ${\dot
  H}(t_m) \ne 0$ and 
$$ \phi_{RS}^2(t_m) = \frac{3}{4 \pi G (1 + \lambda)}. \eqno(29)$$

Thus, like GR-model in RSII-model too, the scale factor $a(t)$, given by eq.(5a) satisfy eqs.(25a,b) for
$\omega^2 = - 1$ (phantom conformal scalar), but not for $\omega^2
= 1$ (quintessence conformal scalar).

\smallskip
\noindent{\bf Quantum bounce near $t=t_s$}

\smallskip
  
As $H(t_m) = 0,$ eq.(25a) yields

$$\rho_{RS}(t_m) = 0,  \eqno(30a)$$
or
$$ \rho_{RS}(t_m) = - 2 \lambda . \eqno(30b)$$
These results are obtained without making quantum correction in eq.(25a).
Eq.(30a) corresponds to GR-model. Effect of the brane theory is realized in
eq.(30b). So, for RSII model, only eq.(30b) is considered.

Like GR-model, here also quantum correction to eq.(25a) is made replacing 
$\rho_{(RS)}$ by $\rho_{(RS)} + \rho_{A}$ as
$$ H^2 = \frac{8 \pi G}{3} (\rho_{RS}+ \rho_{A}) \Big[ 1 + \frac{(\rho_{RS} + \rho_{A})}{2 \lambda} \Big] \eqno(31)$$

Now using $H(t_m) = 0,$ in eq.(31), it is obtained that
$$ \rho_{RS}(t_m) = - 2 \lambda - \rho_{A}(t_m) = - 2 \lambda + \rho(t_m)  . \eqno(32)$$
where $\rho(t_m)$ is given by eq.(17).

 It shows that, in RSII model,  expansion of the universe stops at a lower energy
density compared to GR case due to brane tension . Later on,like GR case, it contracts during the
interval $t_m < t \le t_s$ to its minimum $a(t_s)$ following $a(t)$,
given by eq.(21a), with 
$B^{\prime}, C^{\prime}$ given by eqs.(21b) and (21c).

Like GR case, here also, at $t=t_s$ quantum correction to eq.(25a) leads to

$$\rho_{s(RS)} = - 2 \lambda + \rho(t_s)  \eqno(33)$$
with $\rho(t_s)$, given by eq.(24a) .

Thus, in this case too, bounce will take place at $t=t_s$ like GR-model. In
\cite{mrs}, bounce  has been discussed using
self-gravitational corrections to Friedmann equation for the brane cosmology,

\bigskip

\noindent \underline{\bf  Gauss-Bonnet model }

\noindent{\bf Classical approach}

\smallskip

Einstein-Gauss-Bonnet theory is a generalization over brane-gravity. So, it is
reasonable to analyze this scenarion in in GB-model also.

Modified Friedmann equations for Einstein- Gauss- Bonnet theory is given as

$$ H^2 = \frac{1}{4 {\tilde \alpha}} \Big[ ( 1 - 4{\tilde \alpha} \mu^2) cosh \Big(\frac{2
  \chi}{3} \Big) - 1 \Big] , \eqno(34)$$
where ${\tilde \alpha} = \frac{1}{8 g_s^2}$ ($g_s$ is the string energy scale), energy
  density $\rho_{(GB)}$ and brane-tension $\lambda_{(GB)}$ is connected to energy parameter
  $\chi$ as
\begin{eqnarray*}
 k_5^2 (\rho_{(GB)} + \lambda_{(GB)}) &=& \Big[\frac{2 ( 1 - 4 {\tilde \alpha} \mu^2)^3}{\tilde \alpha}
\Big]^{1/2} sinh \chi \\ &=&  2 \sqrt{\mu^2 + H^2} [ 3 - 4 {\tilde \alpha} \mu^2 + 8
{\tilde \alpha} H^2]\\ &=& 2 \sqrt{\mu^2 + H^2}[ k_5^2(2 \mu)^{-1} \lambda_{(GB)}^{\rm cri.} + 8
{\tilde \alpha} H^2] ,
\end{eqnarray*}
 $$  \eqno(35a)$$
where  
$$k_5^2 \lambda_{(GB)}^{\rm cri.} = 2 \mu (3 - 4 {\tilde \alpha} \mu^2), \eqno(35b)$$
$$ k_5^2 = \frac{(1 + 4 {\tilde \alpha} \mu^2)}{\mu} k_4^2 = \frac{8 \pi}{M_5^3} \eqno(35c)$$ 
and
$$ k_4^2 = \frac{8 \pi}{M_P^2}. \eqno(35d)$$
$\lambda_{(GB)}^{\rm cri.}$ is obtained using the condition $\rho_{(GB)} = H =
0$ \cite{ms, jf}in eq.(35a).

Like RSII model, here also, matter field is given by the conformal scalar
field $\phi$ used above. So, $\rho_{(GB)}$ and $p_{(GB)}$ are given by eqs.(3c,d).
In GB-model, conservation equation looks like

$$ {\dot \rho_{(GB)}} + 3 H ( \rho_{(GB)} + p_{(GB)} + \lambda_{(GB)} )= 0.
\eqno(36)$$

Connecting eqs.(35a,b,c,d) and (36), it is obtained that
\begin{eqnarray*}
 {\dot H} & =& - \frac{k_4^2 k_5^2 \sqrt{\mu^2 + H^2}}{2 [k_5^2 \mu^2 + 8
    {\tilde \alpha} k_4^2 H^2]} (\rho_{(GB)} + p_{(GB)} + \lambda_{(GB)} ) \\ & =&
    - \frac{k_4^2 k_5^2 \sqrt{\mu^2 + H^2}}{2 [\mu k_5^2 + 8 {\tilde \alpha} k_4^2
    H^2]} \Big[(1 - \frac{\omega^2}{3}) {\dot \phi}^2 + \{\frac{1}{3}(\omega^2
    - 1) {\dot H} + \frac{\omega^2}{3} H^2 \} \phi^2 \\&& + H \phi {\dot \phi}
    +  \lambda_{GB} \Big].
\end{eqnarray*}
$$ \eqno(37)$$

Eqs.(3c) and (33a) yield
$${\dot \phi} = \omega^2 \Big[ - H \phi \pm \sqrt{(1 - \omega^2) H^2 \phi^2 +
  \frac{4 \omega^2}{k_5^2}\sqrt{\mu^2 + H^2} [ 3 - 4 {\tilde \alpha} \mu^2 + 8
{\tilde \alpha} H^2] - 2 \omega^2 \lambda} \Big]. \eqno(38)$$

At $H(t_m) = 0,$ using eqs.(3b,c,d), (35b)  and (38) in eq.(37), it is obtained that
$$\Big[1 + \frac{k_4^2}{6} (\omega^2 - 1) \phi^2 \Big] {\dot H(t_m)} = - \frac{k_4^2}{2}
\Big[ 2(1 - \frac{\omega^2}{3})(\lambda_{GB}^{\rm cri.} - \lambda_{GB}) +
\lambda_{GB} \Big]. \eqno(39)$$

\underline{Case 1. For quintessence conformal scalar $\phi$ i.e. $\omega^2 =
  1$}

In this case, eq.(39) reduces to 
$$ {\dot H(t_m)} = - \frac{k_4^2}{2} \Big[\frac{4}{3}(\lambda_{GB}^{\rm cri.}
- \lambda_{GB}) + \lambda_{GB} \Big]. \eqno(40)$$

It shows that unless $\lambda_{GB} = 4 \lambda_{GB}^{\rm cri.}, {\dot H(t_m)}$
does not vanish, so barring this particular situation, in GB-model, the scale
factor given by eq.(5a) is valid for quintessence conformal scalar too. It is
unlike the case of GR-model as well as RSII-model. 

\underline{Case 2. For phantom conformal scalar $\phi$
  i.e. $\omega^2 = - 1$}

For this case, eq.(39) is obtained as
$$ \Big[1 - \frac{k_4^2}{3} \phi^2(t_m) \Big]{\dot H(t_m)} = - \frac{k_4^2}{2}
\Big[\frac{4}{3}(\lambda_{GB}^{\rm cri.} - \lambda_{GB}) + \lambda_{GB} \Big]. \eqno(41)$$

This equation shows that ${\dot H(t_m)}$ does not exist when $\phi^2(t_m) =
3/k_4^2.$ Moreover ${\dot H(t_m)}$ vanishes when $\lambda_{GB} = (8/5)
\lambda_{GB}^{\rm cri.}$ and $\phi^2(t_m) \ne 3/k_4^2.$ Thus, the scale factor
$a(t)$ given by eq.(5a) is valid for phantom conformal
scalar $\phi$ like GR-model as well as RSII-model, barring the situation when
$\phi^2(t_m) = 3/k_4^2$ and $\lambda_{GB} = (8/5)\lambda_{GB}^{\rm cri.}$

\smallskip
\noindent{\bf Quantum bounce near $t=t_s$}

\smallskip
  
If expansion stops at $t=t_m, H(t_m) = 0,$ so eq.(35a) yields

$$\rho_{GB}(t_m) =  \lambda_{GB}^{\rm cri.} - \lambda_{GB}  . \eqno(42)$$
It corresponds to GR-model in case $\lambda_{GB} = \lambda_{GB}^{\rm cri.}. $
Quantum correction to eq.(35a) modifies the equation (42) as
\begin{eqnarray*}
\rho_{GB}(t_m) &=&   \lambda_{GB}^{\rm cri.} - \lambda_{GB} - \rho_A(t_m) \\
&=&  \lambda_{GB}^{\rm cri.} - \lambda_{GB} + \rho(t_m)  
\end{eqnarray*}
$$ \eqno(43)$$
where $\rho(t_m)$ is given by eq.(17) for GR-model. This equation shows that
$\rho_{\rm GB} (t_m) > \rho(t_m)$, when $\lambda^{\rm cri.}_{\rm GB} >
\lambda_{\rm GB},$ otherwise $\rho_{\rm GB} (t_m) < \rho(t_m)$.

Further universe contracts following the scale factor given by eq.(21a) and
bounces at $t=t_s$. At $t=t_s$, the energy density is obtained as
$$\rho_{GB}(t_s) =  \lambda_{GB}^{\rm cri.} - \lambda_{GB} + \rho(t_s),
\eqno(44)$$
where $\rho(t_s)$ is given by eq.(24a).

\smallskip
\noindent{\bf Conclusion}

\smallskip

In the cosmic picture explored above, it is found that quantum gravity makes
drastic changes in the course of future accelerated universe driven by phantom
dark energy in GR-model and RSII-model. But ,in case of GB-model, similar
results are obtained for both quintessence dark energy and phantom dark energy. It is demonstrated above that, under
quantum effects, universe
expands upto a maximum when energy density grows to $\rho(t_m),\rho_{RS}(t_m) $ , and
$\rho_{GB}(t_m)$, given by eqs.(17), (32) and (43)  in GR , RSII
and GB- models respectively. It
is found that, in RSII  model,due to brane-tension, $\rho_{RS}(t_m)$ is lower than the
same in GR model. In GB-model, $\rho_{GB}(t_m)$ is lower than the
same in GR model if brane-tension is higher than the critical brane-tension,
but for brane-tension lower than the critical brane-tension
$\rho_{GB}(t_m)$ is higher than $\rho(t_m)$.

 Later on, it contracts upto a minimum size $a_s$ at $t=t_s$, when energy density rises to $\rho_s$,
given by eqs.(24a), (33) and (44) for GR , RSII and GB- models respectively. Like $\rho_{RS}(t_m)$,
$\rho_{RS}(t_s)$ is also lower in RSII-model. In GB-model, $\rho_{GB}(t_s)$ is
lower or higher than $\rho(t_s)$ depending on the brane-tension like $\rho_{GB}(t_m)$. Finite energy
density is obtained when $\alpha = \beta = 3$ and $\gamma = 4$ in the scale
factor $a(t)$ given by eq.(18a). It shows that universe will contract during the time
period $t_m < t \le t_s$ and expand for $t >t_s$ exhibiting a bounce at
$t=t_s$. Thus, a possibility of re-birth to an exponentially
expanding non-singular universe for $t > t_s$ is obtained. In the present cosmic
picture, {\em future singularity} is not mild, but cosmic bounce at $t=t_s$
shows its complete avoidance.

\end{document}